\documentclass[11pt]{article}

\newenvironment{numberedlist}
{\begin{list}{\makebox[20pt]{\hss(\arabic{itemno})\enspace}}
             {\usecounter{itemno}\labelwidth 20pt}}{\end{list}}

\newcounter{itemno}

\newcounter{itemno1}

\newcounter{itemno2}

\newcounter{exno}

\newcounter{defno}

\newenvironment{defn}{\refstepcounter{defno}\medskip \noindent {\bf
Definition \thedefno.\ }}{\medskip}

\newcommand{\sep}{\;\vert\;}

\newcommand{\oprove}{\vdash\kern-.6em\lower.7ex\hbox{$\scriptstyle O$}\,}

\newcommand{\Pscr}{{\cal P}}

\newcommand{\pderivation}{{\cal P}\kern -.1em\hbox{\rm -derivation}}
\newcommand{\pderivationl}{{\cal P}\kern -.1em\hbox{\em -derivation}}
\newcommand{\pderivable}{{\cal P}\kern -.1em\hbox{\rm -derivable}}
\newcommand{\pderivablel}{{\cal P}\kern -.1em\hbox{\em -derivable}}
\newcommand{\pderivations}{{\cal P}\kern -.1em\hbox{\rm -derivations}}
\newcommand{\pderivability}{{\cal P}\kern -.1em\hbox{\rm -derivability}}

\newcommand{\all}{\forall}

\newcommand{\ie}{{\em i.e.}}

\newsavebox{\lpartfig}
\newsavebox{\rpartfig}

\newenvironment{exmple}{
 \begingroup \begin{tabbing} \hspace{2em}\= \hspace{3em}\= \hspace{3em}\=
\hspace{3em}\= \hspace{3em}\= \hspace{3em}\= \kill}{
 \end{tabbing}\endgroup}

\newcommand{\lb}{\langle}
\newcommand{\rb}{\rangle}

     {\\* \hspace*{\fill} \end{trivlist}}

\newcommand{\uch}{schoo}
\newcommand{\kch}{\uch}
\newcommand{\sch}{\bigtriangledown}

\newcommand{\exs}{ex_u} 
\newcommand{\exm}{ex_m}
\newcommand{\bc}{bch}
\renewcommand{\Pscr}{D}

\begin{document}
\begin{center}
{\Large {\bf A Logical Approach to Event Handling  in Imperative  Languages}}
\\[20pt] 
{\bf Keehang Kwon}\\
Dept. of Computer  Engineering, DongA University \\
Busan 604-714, Korea\\
  khkwon@dau.ac.kr\\
\end{center}

\noindent {\bf Abstract}: 
 While event handling is a key element in modern interactive programming, it is 
 unfortunate that its theoretical foundation is rather weak. 

To solve this problem, we propose to adopt a game-logical approach  of
computability logic \cite{Jap08} to event handling.  

{\bf keywords:}  event handling, game semantics, interaction, computability logic.


\section{Introduction}\label{sec:intro}

 Event handling is a key element in modern   programming paradigm such as GUI programming.
 Despite  the importance,  modern imperative languages have
 lacked  theoretical foundations  for representing events.
To solve this, we first observe that the problem of representing events reduces to the
problem of representing objects with switching capabilities (by the user).
For example,  an earphone can be switched by the user from being diconnected to being connected.

To represent  objects with switching capabilities,  we propose to adopt a 
sequential-choice-disjunctive
operator in computability logic \cite{Jap08}. To be precise,
 a sequential-choice-disjunctive  statement of the form
$\kch(D_1,\ldots,D_n)$ is allowed in the declarations ($\sch$  was originally used
 in \cite{Jap08}.) where each $D_i$ is a constant declaration or a procedure declaration.
This statement has the
following semantics: Use $D_1$ first. If the user types $Esc$, then switch to use $D_2$.
  For example,
an  earphone, declared as $\kch(on == 0, on == 1)$, indicates that it
is originally disconnected. However, if the user types $Esc$, then it switches its status to
being connected. Hence, it provides a form of {\em dynamic} knowledgebases\cite{Jap08}.

 On the other hand, the use of $\kch$ in the main program requests the machine to sequentially choose one
among several alternatives. Therefore, it is identical to the old $if$-$then$-$else$ statement.

\section{The Language}\label{sec:logic}

The language is   core C 
 with  procedure definitions. It is described
by $G$-,$C$- and $D$-formulas given by the syntax rules below:
\begin{exmple}
\>$G ::=$ \>  $\top \sep print(x) \sep A \sep cond \sep x = E \sep  G;G \sep \kch(G_1,\ldots,G_n)$ \\
\>$C ::=$ \>  $c == E \sep A = G\ \sep \all x\ C $\\
\>$D ::=$ \>  $C \sep D \land D \sep  \kch(D,\ldots,D) $\\
\end{exmple}
\noindent
 Here, $\top$ is a true statement, $A$  represents a head of an atomic procedure  of the form $p(x_1,\ldots,x_n)$,
$x = E$ is an assigment statement and $cond$ is a boolean condition. Note that
a boolean condition is a legal statement in this language. 
$c == E$  is a constant  declaration with value $E$.

In the sequel, $G$-formulas will function as the
main  statement, and a  $D$-formula  will constitute  a program.
$\theta$ represents the substitution state which is a set of variable-value bindings.
Note that $\theta$ is initially set to an empty set and will be updated  during execution via the assignment statements.

We need some definitions first. We understand a formula $K \supset H$ as $\neg K\ \lor H$ and 
a procedure declaration $A = G$ as $G \supset A$.
Now an {\em elementarization} of a formula $F$ is obtained by

\begin{itemize}

\item replacing in $F$ all the surface occurrences of $\kch(G_0,\ldots,G_n)$ by $G_0$, and

\item replacing in $F$ all the surface occurrences of $\kch(D_0,\ldots,D_n)$ by $D_0$, and

\item replacing in $F$ all the assignment statements by $\bot$.

\item replacing in $F$ all the {\em print} statements by $\bot$.

\item replacing in $F$ all the occurrences of $G_0;G_1$  by $G_0\land G_1$

\end{itemize}
\noindent
A formula is said to be {\em stable} if its elementarization is classically valid.

Given $\Pscr$ and $G$, we assume that the relation $stable$ which does the following is available.

\begin{itemize}

\item $stable(\Pscr,G,0)$ if $\Pscr\supset G$ is instable 
and  a move is available
for the machine.

\item $stable(\Pscr,G,-1)$ if $\Pscr\supset G$ is instable and  no moves are available  
for the machine

\item $stable(\Pscr,G,1)$ if $\Pscr\supset G$ is stable and a move is available
for the user.

\item $stable(\Pscr,G,2)$ if $\Pscr\supset G$ is stable and  no moves are available  
for the user.

\end{itemize}

We will  present an interpreter which is adapted from \cite{Jap08}.
The basic idea is for the execution to avoid  backtrackings because backtracking is $not$
acceptable in interactive programming. 

Note that this interpreter  alternates between 
 the machine move phase 
and the user move phase.  
In  the machine move phase (denoted by $\exm$), the machine tries to make a move by
executing  the assignment statement, the print statement or the $\kch$ statement.
 In the user move mode (denoted by $\exs$), the user tries 
to make a move and produce a new  program $\Pscr'$
by executing the $\kch$ declarations.

Below, we need some notations at the meta level:
$S$\ sand\ $R$ denotes the  sequential execution of two tasks. 
$S$\ pand\ $R$ denotes the  parallel conjunctive execution of two tasks.
$S$\ por\ $R$ denotes the  parallel disjunctive execution of two tasks.
$S$\ choose\ $R$ denotes the  selection between two tasks. 

Below, the notation $S \leftarrow R$ denotes  reverse implication, \ie, $R \rightarrow S$ at the meta level.

\begin{defn}\label{def:semantics}
Let $G$ be a main statement, let $\Pscr$ be a program and let $\theta$ be a substitution.
Then the notion of   executing $\lb \Pscr,G,\theta\rb$ successfully and producing a new 
substitution $\theta_1$-- $ex(\Pscr,G,\theta,\theta_1)$ --
 is defined as follows: \\

 $ex(\Pscr, G,\theta,\theta_1)\ \leftarrow$ 
\begin{exmple}
   $stable(\Pscr, G,I)$  sand  \% check first whether the execution is stable \\
              choose( \\
\>            $I = 1$ sand $read(EV)$ sand $\exs(\Pscr,EV,\Pscr_1)$ sand $exec(\Pscr_1, G,\theta,\theta_1)$, \%  user's move.\\
\>            $I = 2$,  \% nothing for the user to do. Execution succeeds. \\
\>               $I = 0$ sand  $\exm(\Pscr,G,\theta,G',\theta')$ sand 
    $ex(\Pscr, G',\theta',\theta_1)$,  \%   machine's move     \\
\>              $I = -1$  \% nothing for the machine to do.  Execution  fails. \\
\>             ) \\
\end{exmple}
\noindent where $\exm(\Pscr,G,\theta,G_1,\theta_1)$ is defined as follows:
\end{defn}

Note that this $\exm$ phase  alternates between two subphases:
 the backchaining subphase 
and the goal reduction subphase until the machine makes a move.

\begin{defn}\label{def:semantics}
Let $G$ be a main statement, let $\Pscr$ be a program and let
$\theta$ be a substitution.
Then the notion of the machine making a single move in  $(\Pscr,G,\theta)$  and producing a new
substitution $\theta_1$, and a new goal $G_1$ -- $\exm(\Pscr,G,\theta,G_1,\theta_1)$ --
 is defined as follows:
\begin{numberedlist}

\item    $\bc((A = G_1),\Pscr,A,\theta,G,\theta_1) \leftarrow$
  $\exm(\Pscr,G_1,\theta,G,\theta_1)$ \% A matching procedure for $A$ is found.

\item    $\bc(\all x D,\Pscr,A,\theta,G,\theta_1)\ \leftarrow$    $\bc([t/x]D,
\Pscr, A,\theta,G,\theta_1)$. \% argument passing

\item    $\bc(D_0 \land D_1,\Pscr,A,\theta,G,\theta_1)\ \leftarrow$    $\bc(D_0,
\Pscr, A,\theta,G,\theta_1)$ por $\bc(D_1, \Pscr, A,\theta,G,\theta_1)$                      

\item    $\bc(\kch(D_0,\ldots,D_n),\Pscr,A,\theta,G,\theta_1)\ \leftarrow$    $\bc(D_0,
\Pscr, A,\theta,G,\theta_1)$ \% only the first one is currently active.       

\item    $\exm(\Pscr,A,\theta,G,\theta_1)\ \leftarrow$   
  $\bc(\Pscr,\Pscr, A,\theta,G,\theta_1)$. \% a procedure call

\item  $\exm(\Pscr,\top,\theta,\Pscr,\top,\theta)$. \% True is always a success.

\item  $\exm(\Pscr,cond,\theta,\Pscr,cond,\theta)\ \leftarrow$     $eval(\Pscr,\theta,cond)$. \% a boolean condition

\item  $\exm(\Pscr,x=E,\theta,\Pscr,\top,\theta\uplus \{ \lb x,E' \rb \})\ \leftarrow$  $eval(\Pscr,E,E')$.
\% the assignment statement. Here, 
$\uplus$ denotes a set union but $\lb x,V\rb$ in $\theta$ will be replaced by $\lb x,E' \rb$.

\item  $\exm(\Pscr,print(x),\theta,\Pscr,\top,\theta)\ \leftarrow$  
print  $\theta(x)$.
\% the print statement. 

\item  $\exm(\Pscr,G_0; G_1,\theta,G,\theta')\ \leftarrow$  \\
 $\exm(\Pscr,G_0,\theta,G'_0,\theta_0)$  sand \% try $G_0$ first. \\
choose( \\
 $G_0 \neq  G'_0$ sand  $G = G'_0; G_1$ sand $\theta' = \theta_0$, \%   a move is made in $G_0$ \\
  $G_0 =  G'_0$ sand $\exm(\Pscr,G_1,\theta,G'_1,\theta_1)$ sand $G = G_0; G'_1$ sand \\
$\theta' = \theta_1$ \% try $G_1$ \\
)

\item $\exm(\Pscr,\kch(G_1,\ldots,G_n),\theta,\kch(G_2,\ldots,G_n),\theta)$. \% make a 
switch.

\end{numberedlist}
\end{defn}
\noindent 

The procedure  $\exs(\Pscr,EV,\Pscr')$ is defined as follows:

\begin{defn}\label{def:semantics}
Let $\Pscr$ be a program.
Then the notion of the user making a  move in  $D$ using $EV$  and producing a new
program $\Pscr'$ -- $\exs(\Pscr,EV,\Pscr')$ --
 is defined as follows:

\begin{numberedlist}

\item $\exs(C,w.Esc,C)$.

 \item $\exs(D_0\land D_1,w.Esc, D'_0 \land D_1)  \leftarrow$   
$w$ is the address of $D_0$ and $\exs(D_0,w'.Esc, D'_0)$. 
Here $w'$ is the  location adjusted from $w$.

 \item $\exs(D_0\land D_1,w.Esc, D_0 \land D'_1)  \leftarrow$   
$w$ is the address of $D_1$ and $\exs(D_1,w'.Esc, D'_1)$. 
Here $w'$ is the  location adjusted from $w$.

\item $\exs(\kch(D_0,\ldots,D_n),w.Esc, \kch(D_1,\ldots,D_n))  \leftarrow$ \\
$w$ is the address of $\kch$. \% event processed \\

\end{numberedlist}
\end{defn}
\noindent

\section{Examples }\label{sec:modules}

The following  code prints the price of a car, based on the user's
choice of a model.

\begin{exmple}
\% constant declaration \\
print(``type Esc to switch''); \\
 \kch(model == BMW320, model == BMW520, model == BMW740) 
\end{exmple}

\noindent with the following $G$ formula:

\begin{exmple}
        \kch( \\
 \>            model == BMW320;   price = \$32,000; print(price), \\
  \>           model == BMW520;   price  = \$54,000; print(price),  \\
 \>            model == BMW740;  price = \$82,200; print(price))
\end{exmple}

\noindent 
 Initially, the execution is instable. Therefore, the 
machine executes the first three statements and prints     \$32,000.
 As the execution becomes
stable, the machine  waits for the user to make a move (by
 switching to the second model). If the user did switch to the BMW520 by typing ESC, then the execution becomes instable and the machine also 
switches to the second one and so on. If the user makes no move, the machine keeps waiting.


\section{Conclusion}\label{sec:conc}

So far, we have extended  the basic  C with the addition of
$\kch$ statements. 
These statements can be used in the declarations or in the main program and 
are  useful for representing event handling.

Event handling is  a very challenging subject, especially in the presence of asynchrous events.
Note that we have dealt with  simple synchronous events.
In the future, we hope to include asynchronous events as well.



\bibliographystyle{plain}


\end{document}